\documentstyle[12pt]{article} 
\newcommand{\oB}{\vert_{\partial M}=0}

\begin{document} 

\def\bce{\begin{center}} 
\def\ece{\end{center}} 
\def\beq{\begin{eqnarray}} 
\def\eeq{\end{eqnarray}} 
\def\ben{\begin{enumerate}} 
\def\een{\end{enumerate}} 
\def\ul{\underline} 
\def\ni{\noindent} 
\def\nn{\nonumber} 
\def\bs{\bigskip} 
\def\ms{\medskip} 
\def\wt{\widetilde} 
\def\wh{\widehat} 
\def\dsp{\displaystyle} 
\def\brr{\begin{array}} 
\def\err{\end{array}} 
 
 
\hfill IEEC 98-91 (revised)
 
\hfill hep-th/9809211 
 
\hfill December 1998 
 
\vspace*{15mm}

\begin{center} 
 
{\LARGE \bf Heat Kernel Coefficients for Chern-Simons Boundary Conditions in 
QED}

\vspace{12mm} 
 
\medskip

{\sc E. Elizalde}\footnote{E-mail address: 
 elizalde@ieec.fcr.es, eli@zeta.ecm.ub.es}\\ 
Consejo Superior de Investigaciones Cient\'{\i}ficas (CSIC),\\ 
Institut d'Estudis Espacials de Catalunya (IEEC), \\ 
Edifici Nexus 201, Gran Capit\`a 2-4, 08034 Barcelona, Spain, \\ and \ 
Departament ECM and IFAE, Facultat de F\'{\i}sica, \\ 
Universitat de Barcelona, Diagonal 647, 
08028 Barcelona, Spain,  \\ 
and \\ 
{\sc D.V. Vassilevich}\footnote{E-mail address: 
Dmitri.Vassilevich@itp.uni-leipzig.de . On leave from: 
Department of Theoretical Physics, St. Petersburg University, 
198904 St. Petersburg, Russia} \\ 
Universit\"at Leipzig, Fakult\"at f\"ur Physik und Geowissenschaften \\ 
Institut f\"ur Theoretische Physik\\ 
Augustusplatz 10/11, 04109 Leipzig, Germany

\vspace{20mm} 
 
{\bf Abstract} 
 
\end{center} 

We consider the four dimensional Euclidean Maxwell theory with
a Chern--Simons term on the boundary. The corresponding gauge
invariant boundary conditions become dependent on tangential
derivatives. Taking the four-sphere as a particular example,
we calculate explicitly a number of the first heat kernel coefficients
and obtain the general formulas that yields any desired coefficient. 
A remarkable observation is that the coefficient $a_2$, which defines the
one--loop counterterm and the conformal anomaly, does not
depend on the Chern--Simons coupling constant, while the
heat kernel itself becomes singular at a certain (critical)
value of the coupling. This could be a reflection of a 
general property of Chern--Simons theories.

\vfill 
 
 
 
 
\newpage 
 
 
\section{Introduction} 

The heat kernel expansion plays an important role in quantum
field theory. In fact, the
ultraviolet divergences and renormalization structure
of the theory at the one--loop approximation are encoded
in a few of the first heat kernel coefficients $a_n$. For massless theories
in $d$ dimensions the only divergent term is proportional
to $a_{d/2}$. The heat kernel expansion is closely related to
the quantum anomalies, most notably, to the conformal and chiral
anomalies. It  has also numerous applications in mathematics.
No wonder that much effort has been devoted to calculate
the heat kernel expansion. For operators of the Laplace type,
the coefficients $a_n$ are known fairly well, both for manifolds
without a boundary and for Dirichlet and Neumann boundary
conditions in manifolds with a boundary.
More details on the physical applications of the
heat kernel expansion can be found in the recent monographs
\cite{phys}. The mathematical background is described in \cite{Gilkey}.

During the last couple of years, considerable interest has
been attracted by the  
heat kernel expansion for  boundary conditions depending on 
tangential derivatives \cite{AE-1}--\cite{DK2}, following 
the pioneering work \cite{MO}. In the mathematical literature such 
boundary conditions are called oblique. For physicists, the interest towards 
oblique boundary conditions has been primarily motivated by problems 
of quantum gravity \cite{EKbook}. It has been realized that the 
diffeomorphism 
invariant boundary conditions for the graviton inevitably contain  
tangential derivatives \cite{grav,AE-1,MoSi,DV-2d}. Oblique boundary 
conditions may also appear in electrodynamics, when the properties of a 
physical boundary depend on the photon frequency. A more formal 
motivation for studying oblique boundary conditions can be found 
in \cite{BGV2}, where tangential derivatives arise when one calculates 
vacuum expectation values of second order differential operators. 
 
The usual strategy  \cite{AE-1}--\cite{DK2} employed in the calculation of 
the heat kernel coefficients for oblique boundary conditions is 
as follows. By means of the corresponding invariant theory,  
each coefficient 
can be expressed through several universal functions, which are 
calculated by using ordinary conformal variation techniques and some 
explicit examples. However, up to now, only a very few examples admitting 
calculation of the spectrum and an explicit evaluation of the heat 
kernel were known \cite{AE-1,DK1,DK2}. In the present paper we 
study the Euclidean Maxwell theory with a Chern--Simons boundary 
term, which generates oblique boundary conditions. We find the  
spectrum of the Laplace operator on a ball and calculate the 
heat kernel expansion. 
 
Apart from its application to the theory of the heat kernel expansion, 
our model may be interesting by itself. Tangential derivatives 
appear on the boundary conditions in a very natural way. An intriguing 
property of the model, established in the present paper, is that the 
conformal anomaly and the one--loop counterterm do not depend on the 
charge standing in front of the Chern--Simons action.
Other heat kernel coefficients become singular for a certain
value of this charge, what suggests the existence of a critical
value of the Chern--Simons charge. Let us recall that
the Chern--Simons gauge theory introduced in \cite{ChSG}
exhibits some unusual and intriguing properties, as,
for example, generation of states with fractional statistics
first observed by Wilczek \cite{Wilczek}. Later on this phenomenon
was used in the theory of the Fractional Quantum Hall Effect (FQHE) (for
a recent review see \cite{FQHE}). In spite of the fact that
 the geometry of our model is different from the ones
usually considered in the FQHE, the results of our calculation
may be important in order 
to gain insight into the general properties of Chern--Simons
theories, as e.g. non-renormalization theorems \cite{Ovch}.
 
The paper is organized as follows. In  section 2 we formulate 
the model, discuss its general properties and define the eigenfunctions 
on the Euclidean ball. In section 3 we calculate the heat kernel 
asymptotics explicitly. A  discussion on the possible significance
of the results obtained 
(a list of the first ten heat kernel coefficients is given in Table 1) 
 is presented in section  4. 
  
\section{Chern--Simons boundary conditions} 
Consider the action for the Euclidean Maxwell theory on a manifold 
$M$ 
\begin{equation} 
S=\frac 14 \int_M d^4x\ g^{\frac 12} F_{\mu\nu}F^{\mu\nu} , 
\quad F_{\mu\nu}=\partial_\mu A_\nu -\partial_\nu A_\mu. 
\label{act} 
\end{equation} 
We can add to the action (\ref{act}) a boundary term. If we require 
gauge and coordinate invariance to hold and, if we do not want to introduce 
any dimensional parameter, the only choice available 
 is the Chern--Simons action 
\begin{equation} 
S_{CS}=\frac a2 \int_{\partial M} d^3x \ \varepsilon^{ijk} A_i 
\partial_j A_k , 
\label{SCS} 
\end{equation} 
where $\varepsilon^{ijk}$ is the Levi-Civita tensor and the $x^j,\ j=1,2,3 
$, are coordinates on the boundary $\partial M$, $a$ being a real 
parameter. 
 
To calculate the path integral, it is convenient to write $S+S_{CS}$ in the 
form $\int ALA$, with $L$ a second order differential operator. To this end, we 
integrate by parts, obtaining 
\begin{eqnarray} 
S+S_{CS}&=&\frac 12 \int_M d^4x\ g^{\frac 12} 
A_\mu (-g^{\mu\nu}\Delta +\nabla^\nu \nabla^\mu )A_\nu \nonumber \\ 
\ &\ & +\frac 12 \int_{\partial M} d^3x 
(h^{\frac 12}(\partial_NA_i-\partial_iA_N)A_i+ a\varepsilon^{ijk} A_i 
\partial_j A_k) , 
\label{Stot} 
\end{eqnarray} 
where $\Delta$ is the Laplace operator, $N$  the outward pointing 
normal vector, and $h$ is the determinant of the induced metric 
on $\partial M$. 
There are, at least, two sets of gauge invariant boundary 
conditions which ensure vanishing of the surface term in (\ref{Stot}). 
The first set is ordinary relative (or magnetic) boundary conditions: 
$A_i\oB$, $i=1,2,3$, $(\partial_N+k)A_N\oB$. Here $k$ is the 
trace of the second fundamental 
form of the boundary. These boundary conditions have been extensively 
studied in the literature, but we shall not consider them here. Another 
possible set is the following: 
\begin{equation} 
A_N\oB ,\quad 
(\partial_NA_i+ah^{-\frac 12}{\varepsilon_i}^{jk}\partial_jA_k) 
\oB \ \ i=1,2,3 . 
\label{bcon} 
\end{equation} 
For $a=0$, Eqs. (\ref{bcon}) become ordinary absolute (or electric) 
boundary conditions. 
 
The boundary conditions (\ref{bcon}) possess two properties 
which make them useful for quantum electrodynamics. First, 
they are gauge invariant. This means that if  $A_\mu$ satisfies 
(\ref{bcon}), then $A+\partial \phi$ also satisfies (\ref{bcon}), 
provided that 
\begin{equation} 
\partial_N\phi\oB. \label{bcg} 
\end{equation} 
After quantization, Eq. (\ref{bcg}) becomes the boundary condition for the 
ghost field. Second, the Laplace operator is symmetric, i.e., 
\begin{equation} 
\int_M d^4x\ g^{\frac 12} (A^{(1)\mu}\Delta A_\mu^{(2)}- 
A^{(2)\mu}\Delta A_\mu^{(1)})=0, 
\label{symm} 
\end{equation} 
if both $A^{(1)}$ and $A^{(2)}$ satisfy (\ref{bcon}). 
 
It is natural to impose the Lorentz gauge condition 
\begin{equation} 
\nabla^\mu A_\mu =0. \label{gcon} 
\end{equation} 
In this case the path integral is given by 
\begin{equation} 
Z={\det }^{-\frac 12}_{T}(-\Delta ) 
\times {\det }_S^{\frac 12} (-\Delta ), 
\label{pint} 
\end{equation} 
where the first determinant is calculated on the space of transversal 
vectors, with the boundary condition (\ref{bcon}), and the second one 
on the space of scalar fields satisfying (\ref{bcg}). 
 
As an example, consider a ball with the metric 
\begin{equation} 
ds^2=(dx^0)^2+(x^0)^2d\Omega^2, \quad 0\le x^0 \le r , \label{met} 
\end{equation} 
where $d\Omega^2$ is the metric on the unit sphere $S^3$. 
 
We can use the basis of Ref. \cite{vjmp1} in the space of transversal 
vector fields: $\{ A^T\} =\{ A^\perp , A(\psi )\}$, where 
\begin{eqnarray} 
A^{\perp}_0&=&0, \quad \ ^{(3)}\nabla^iA^{\perp}_i=0, \nonumber \\ 
A_0(\psi )&=&\ ^{(3)}\Delta r \psi , \quad 
A_i (\psi )=-^{(3)}\nabla_i(\partial_0+\frac 1r ) 
r \psi , \ \ i=1,2,3, \label{bas} 
\end{eqnarray} 
$\psi$ being a scalar field, and $\ ^{(3)}\nabla$ and $\ ^{(3)}\Delta$ 
the covariant derivative and the Laplacian on $S^3$, respectively.  
 
The boundary conditions for the field $\psi$ do not depend 
on $a$. Hence, the contribution of $\psi$ to the heat kernel 
and functional determinant is the same as for the absolute 
boundary condition \cite{vjmp1,elv2}. Therefore, let us 
concentrate on the $A^\perp$ contribution. The operator 
$h^{-\frac 12}\varepsilon^{ijk}\partial_j$ can be diagonalized 
on the unit sphere $S^3$ (see, e.g., \cite{vlmp}). It has for eigenvalues 
$\pm (l+1)$, with degeneracies given by 
\begin{equation} 
D_l^\pm = l(l+2),\quad l=1,2,\dots . \label{deg} 
\end{equation} 
The corresponding eigenvalues of the Hodge--de~Rham Laplacian on $S^3$ are 
$-(l+1)^2$. The eigenfunctions of the vector Laplace operator on the unit 
ball are found to be 
\begin{equation} 
J_{l+1} (\lambda x^0) Y^{\pm}_l (x^i), 
\label{eigen} 
\end{equation} 
where $J_{l+1}$ are ordinary Bessel functions and $Y^{\pm}_l$ are vector 
spherical harmonics. 
The eigenvalues of the Laplacian are $-\lambda^2$, where the $\lambda$'s 
are defined, implicitly, through the equation for the boundary condition: 
\begin{equation} 
\lambda^\pm_l {J'}_{l+1}(\lambda^\pm_l )\pm  
\frac {a(l+1)}r J_{l+1}(\lambda^\pm_l )=0. 
\label{condi} 
\end{equation} 
The prime denotes here differentiation with respect to the argument. 
The contribution of $A^\perp$ to the path integral reads 
\begin{equation} 
Z^\perp =\prod_{\lambda^\pm_l} (\lambda_l^-\lambda_l^+)^{l(l+2)}. 
\label{contr} 
\end{equation} 
If one proceeds, as usually, by splitting the calculation of the 
determinant into two parts, one has to take into account a possibly 
non-vanishing determinant anomaly (see, e.g., \cite{evz}).

\section{The zeta function: calculation of the singularities} 

In the following, we will use the 
connection between the zeta function and the heat kernel: 
\beq 
\zeta^\pm (s) =\frac 1 {\Gamma (s)} \int\limits_0^{\infty} dt\,\, 
t^{s-1} \sum_k e^{-(\lambda^\pm_k)^2 t}, \label{n6} 
\eeq 
in order to obtain the heat kernel coefficients $C_j^\pm$ or 
$a_n^\pm$: 
\begin{eqnarray} 
&&\sum_k e^{-(\lambda^\pm_k)^2 t}=\frac 1{(4\pi )^2} 
\sum_{n=0}^\infty t^{\frac n2 -2} a_{n/2}^\pm \ ,\nonumber \\ 
&&\frac 1{(4\pi )^2} a_{n/2}^\pm = 
\frac 1{(4\pi )^{3/2}} C_{(n-1)/2}^\pm = 
\mbox{Res}\, \left[ \zeta^\pm (s)\Gamma (s)\right]_{s=2-\frac n2}. 
\label{defan} 
\end{eqnarray} 
The shifted heat kernel coefficients $C_j^{tot}=C_j^+ +C_j^-$ 
 are commonly used in Casimir energy calculations.  
Calling $\Phi_l^\pm$ the functions that determine implicitly 
the spectral values corresponding to the given boundary conditions 
(\ref{condi}), the zeta function is obtained as 
\beq 
\zeta^\pm (s) = \sum_{l=0}^\infty D^\pm_l (\lambda^\pm_l)^{-2s} = 
 \sum_{l=0}^\infty l(l+2)\int_\gamma \frac{dk}{2\pi i} k^{-2s} 
\frac{\partial}{\partial k} \ln \Phi_l^\pm (kr), 
\eeq 
with 
\beq 
 \Phi_l^\pm (kr) \equiv {J'}_{l+1}(kr) \pm \frac{a(l+1)}{kr} J_{l+1}(kr), 
\eeq 
where we have transformed the
sum over the spectrum into a contour integral which goes around all the
real zeros of the function $\Phi$ defining the boundary
conditions. 

By doing this we have reduced the problem to the evaluation
of certain contour integrals, which can be solved in an
explicit ---though technically quite involved--- way. The interested 
reader is addresed to  the Refs. \cite{bek1,ebk1}, where this powerful method
is described in great detail ---in the particular case $a=0$.
Having understood this case, however, 
the situation here is not difficult to grasp,
the main differences being in Eqs. (\ref{e21}) and (\ref{e22}) to follow,
which will lead, in the end, to more involved hypergeometric functions. In
any case, the final expressions will be still explicit, allowing for 
the straightforward use of very quick algebraic computation machinery,
to obtain the final expressions of the heat kernel coefficients to
 {\it any} order. This
will demonstrate, once more, the power (and adaptability) of the 
zeta-function procedure to perform these sort of calculations
(which are extremely cumbersome, by any means). 
 
By deforming the integration contour around  the imaginary axis, 
 in the usual way \cite{bek1,ebk1}, we obtain 
\beq 
\zeta^\pm (s)& =& \frac{\sin (\pi s)}{\pi} \sum_{l=0}^\infty l(l+2) 
 \int_0^{\infty} dk\, k^{-2s} \nn \\ && \times \, 
 \frac{\partial}{\partial k} \ln \left[k^{-(l+1)} 
\left( {I'}_{l+1}(kr) \pm \frac{a(l+1)}{kr} I_{l+1}(kr)\right)\right]. 
\eeq 
Introducing into the equation the asymptotic expansions corresponding to 
$I$ and $I'$ when $k,l\to \infty$, with $z=kr/(l+1)$ fixed, we can write 
the 
 zeta function under the form 
\beq 
\zeta^\pm (s)& =& \frac{\sin (\pi s)}{\pi}  \sum_{l=0}^\infty l(l+2) 
\left(\frac{l+1}{r}\right)^{-2s} \int_0^{\infty} dz\, z^{-2s} 
 \frac{\partial}{\partial z} \left\{ (l+1)\, \ln  \left( 
\frac{e^{1/t}}{zt} \sqrt{\frac{1-t}{1+t}} \right)  \right.  \nn \\ 
&& + \left.  \ln \left[ 1+ \sum_{j=1}^\infty \frac{v_j(t)}{(l+1)^j} \pm 
at \left(  1+ \sum_{j=1}^\infty \frac{u_j(t)}{(l+1)^j} \right) \right] 
\right\}, 
\eeq 
where the functions $u_j(t)$ and $v_j(t)$ give the asymptotics of the 
Bessel functions in the way 
\beq 
I_{\nu} (\nu z) &\sim& \frac 1 {\sqrt{2\pi \nu}}\frac{e^{\nu 
\eta}}{(1+z^2)^{\frac 1 4}}\left[1+\sum_{k=1}^{\infty} \frac{u_k (t)} 
{\nu ^k}\right],\nn   \\ 
I'_{\nu} (\nu z) &\sim& \frac 1 {\sqrt{2\pi \nu}}\frac{e^{\nu 
\eta}(1+z^2)^{\frac 1 4}} z \left[1+\sum_{k=1}^{\infty} \frac{v_k (t)} 
{\nu ^k}\right], 
\eeq 
with $t=1/\sqrt{1+z^2}$ and $\eta =\sqrt{1+z^2}+\ln 
[z/(1+\sqrt{1+z^2})]$. 
The first few coefficients are given in \cite{as1}, 
higher coefficients are easy to obtain  using the recursions 
\beq 
u_{k+1} (t) &=& \frac 1 2 t^2 (1-t^2) u'_k (t) +\frac 1 8 \int\limits_0^t 
d\tau\,\, (1-5\tau^2 ) u_k (\tau ),\nn \\ 
v_k (t) &=& u_k (t) +t(t^2-1) \left[ \frac 1 2 u_{k-1}(t) 
+tu'_{k-1}(t)\right] 
\eeq 
(see  \cite{bek1,ebk1}). It is convenient to write 
 the last logarithmic function in the 
expression of $\zeta^\pm (s)$ 
above as a series 
\beq 
 \ln \left[ 1+ \sum_{j=1}^\infty \frac{v_j(t)}{(l+1)^j} \pm 
at \left(  1+ \sum_{j=1}^\infty \frac{u_j(t)}{(l+1)^j} \right) \right] 
&=& \ln (1\pm at)+ \sum_{n=1}^\infty F_n^\pm (t) \, (l+1)^{-n}, 
\label{e21}\\   
F_n^\pm (t) &=&  \frac{p_{4n}^\pm (t)}{(1\pm at)^n},\nn 
\eeq 
with $p_{4n}^\pm (t)$ a polynomial in $t$ of degree $4n$, which has 
the form 
\beq 
p_{4n}^\pm (t) = \sum_{j=0}^{3n} z_{n,j}^\pm t^{n+j}. \label{e22}
\eeq 
The use of this expansion can be based on the corresponding one 
for the case $a=0$ (see Refs. \cite{bek1,ebk1}) and will be fully justified
by the subsequent calculations. For any heat kernel coefficient
only a limited number of terms in (\ref{e21}) are needed.            
The integrations are readily performed using the basic result (Re $s < 1$): 
\beq 
\int_{0}^\infty dz \, z^{-2s} \, \frac{\partial}{\partial z} 
\left[ \frac{t^k}{(1 \pm at)^n} \right] & =& 
- \frac{\Gamma (1-s)}{\Gamma (n)} \sum_{h=0}^\infty (\mp a)^h 
\frac{\Gamma (n+h)\Gamma \left( s + \frac{k+h}{2} \right)}{ 
h!\  \Gamma \left( \frac{k+h}{2} \right)} \nn  \\ &&  \hspace*{-4cm} = 
\Gamma (1-s) \left[  \pm a \, n \, \frac{\Gamma  \left( s + 
\frac{k+1}{2} \right)}{ \Gamma  \left(\frac{k+1}{2} \right)} 
H \left(\left\{ \frac{n+1}{2},  \frac{n}{2}+1, s+\frac{k+1}{2} \right\}, 
\left\{ \frac{3}{2},\frac{k+1}{2}  \right\},  a^2  \right) \right. \nn \\ 
&& - \left. \frac{\Gamma  \left( s + 
\frac{k}{2} \right)}{ \Gamma  \left(\frac{k}{2} \right)} 
H \left(\left\{ \frac{n+1}{2},  \frac{n}{2}, s+\frac{k}{2} \right\}, 
\left\{ \frac{1}{2},\frac{k}{2}  \right\},  a^2  \right) \right],  
 \label{23} \eeq 
where $H$ is the generalized hypergeometric function. 
Observe that for the sum  over the two signs of $a$ the result 
simplifies to twice the second term on the r.h.s. of Eq. (\ref{23}). 
 
However, one has to take into account that, as it stands,  $\zeta^\pm 
(s)$ is a divergent quantity. 
We shall extract the singular parts of the zeta function 
according to the prescription in  \cite{bek1,ebk1}, 
namely, by splitting the zeta function into a regular and 
a singular part and adding and subtracting a number, say $M$, of 
leading terms of the assymptotic expansion: 
\beq 
\zeta^\pm (s) = Z^\pm (s) + \sum_{i=-1}^M A_{i,a}^\pm (s). 
\eeq

The functions $A^\pm_{i,a}(s)$ are found to be: 
\beq 
A_{-1}^\pm (s) &=& \frac{r^{2s}\Gamma (s-1/2)}{4\sqrt{\pi} \,\Gamma (s+1)} 
\left[ \zeta (2s-3) - \zeta (2s-1) \right], \nn \\ 
A_0^\pm (s) &=& \frac{r^{2s}}{4} 
\left[ \zeta (2s-2) - \zeta (2s) \right],  \nn \\ 
A_{0,a}^\pm (s) &=&\frac{r^{2s}}{2\, \Gamma (s)} 
\left[ \zeta (2s-2) - \zeta (2s) \right] 
\sum_{h=1}^\infty (\mp a)^h \frac{\Gamma (s+h/2)}{\Gamma (1+h/2)}, 
\label{As} \\ A_{n,a}^\pm (s) &=&- \frac{r^{2s}}{ \Gamma (s)} 
\left[ \zeta (2s+n-2) - \zeta (2s+n) \right] \nn \\ &&  \times 
\sum_{j=0}^{3n}   
\frac{z_{n,j}^\pm }{\Gamma (n)}\sum_{h=0}^\infty 
(\mp a)^h \frac{\Gamma (n+h) \Gamma (s+\frac{j+n+h}{2})}{h! \, 
\Gamma (\frac{j+n+h}{2})} ,\nn 
\eeq 
where $\zeta (s)$ is the Riemann zeta function. 
Notice that all the series that appear here are in fact of the form
corresponding to Eq. (\ref{23}) and thus give rise to 
generalized hypergeometric functions (the result simplifies, 
when adding the $\pm$ contributions).
Let us study the poles of $A_n^\pm$ at half integer numbers, $s=3/2-k, 
\ k\in \, \mbox{\bf N}$. 
For even $n=2p$ the poles of the gamma functions at $s=3/2-k$ 
on the r.h.s. of (\ref{As}) do not contribute to $A^{tot}=A^++A^-$ 
due to the property $z^+_{n,j}=(-1)^jz^-_{n,j}$. For odd $n=2p-1$ 
the Riemann zeta functions on the r.h.s. have poles at integer 
values of $s$ only. 
{}From those expressions, the following residua are obtained ($k \in \, 
\mbox{\bf N}$): 
\beq 
&&\hspace{-6mm}\mbox{Res}\, A_{-1}^{tot} (1/2) = \frac{r}{2\, \pi}, \nn \\ 
&&\hspace{-6mm}\mbox{Res}\, A_{0,a}^{tot} (3/2) = 
 \frac{ r^3}{2} \left[(1-a^2)^{-3/2} -1\right], \nn \\ 
&&\hspace{-6mm}\mbox{Res}\, A_{0,a}^{tot} (1/2) = 
 -\frac{ r}{2} \left[(1-a^2)^{-1/2} -1\right], \nn \\  
&&\hspace{-6mm}\mbox{Res}\, A_{2n,a}^{tot} (3/2-n) = 
- \frac{r^{3-2n}}{\Gamma (3/2 -n)} \left\{ 
\sum_{j=0}^{3n}\frac{z_{2n,2j}^+ }{\Gamma (2n)} \sum_{h=0}^\infty 
 \frac{a^{2h}\Gamma (2n+2h)\,\Gamma (j+h+3/2 )}{(2h)!\, 
\Gamma (n+j+h )} \right. \nn \\ 
&&\hspace{30mm} +\left. \sum_{j=0}^{3n-1} 
\frac{z^-_{2n,2j+1}}{\Gamma (2n)} 
\sum_{h=1}^{\infty} 
\frac{a^{2h-1}\Gamma (2n+2h-1)\, \Gamma (j+h+3/2)}{(2h-1)! 
\Gamma (h+n+j)} \right\} , \ \ n\geq 1, \nn \\ 
&&\hspace{-6mm}\mbox{Res}\, A_{2n,a}^{tot} (1/2-n) = 
 \frac{r^{1-2n}}{\Gamma (1/2 -n)} \left\{ 
\sum_{j=0}^{3n}  \frac{z_{2n,2j}^+ }{\Gamma (2n)} \sum_{h=0}^\infty 
 \frac{a^{2h}\Gamma (2n+2h)\,\Gamma (j+h+1/2 )}{(2h)!\, 
\Gamma (n+j+h )} \right. \nn \\ 
&&\hspace{30mm} +\left. \sum_{j=0}^{3n-1} 
\frac{z^-_{2n,2j+1}}{\Gamma (2n)} 
\sum_{h=1}^{\infty} 
\frac{a^{2h-1}\Gamma (2n+2h-1)\, \Gamma (j+h+1/2)}{(2h-1)! 
\Gamma (h+n+j)} \right\}, \ \ n\geq 1,  \nn \\ 
&&\hspace{-6mm}\mbox{Res}\, A_{2n-1,a}^{tot} (3/2-k) = 
 \frac{(-1)^k 4r^{3-2k}}{\Gamma (3/2-k)} 
\left[ \zeta (2n-2k) -  \zeta (2n-2k+2) \right] \nn \\ && 
\hspace{6mm} \times \, \left\{ 
\sum_{j=0}^{k-n-1}   
\frac{z_{2n-1,2j}^+ }{(2n-2)!} \sum_{h=0}^{k-n-j-1}  
\frac{ (-1)^{n+j+h} (2n+2h-2)!\, 
a^{2h}}{(2h)!\, (k-n-j-h-1)!\, \Gamma (n+j+h-1/2)} \right. \nn \\ 
&& \hspace{-2mm} \left. + \sum_{j=0}^{k-n-2} \frac{ z_{2n-1,2j+1}^+ 
}{(2n-2)!} \sum_{h=1}^{k-n-j-1} \frac{ (-1)^{n+j+h} (2n+2h-2)!\, 
 a^{2h-1}}{(2h-1)!\, (k-n-j-h-1)!\, \Gamma (n+j+h-1/2)} \right\},\nn \\ 
 && \hspace{105mm}  n \leq k-1,\  k\leq 4n,\nn \\ 
&&\hspace{-6mm}\mbox{Res}\, A_{2n-1,a}^{tot} (3/2-k) = 
 \frac{(-1)^k 4 r^{3-2k}}{\Gamma (3/2-k)} 
\left[ \zeta (2n-2k) -  \zeta (2n-2k+2) \right]  \\ && 
\hspace{-2mm} \times \, \left\{ 
\sum_{j=0}^{3n-1}  
\frac{z_{2n-1,2j}^+ }{(2n-2)!} \sum_{h=0}^{3n-j-1}  
\frac{ (-1)^{n+j+h} (2n+2h-2)!\, 
a^{2h}}{(2h)!\, (k-n-j-h-1)!\, \Gamma (n+j+h-1/2)} \right. \nn \\ 
&& \hspace{6mm} \left. + \sum_{j=0}^{3n-2} \frac{ z_{2n-1,2j+1}^+ 
}{(2n-2)!} \sum_{h=1}^{3n-j-1} \frac{ (-1)^{n+j+h} (2n+2h-2)!\, 
 a^{2h-1}}{(2h-1)!\, (k-n-j-h-1)!\, \Gamma (n+j+h-1/2)} \right\} 
, \nn \\ && \hspace{105mm} n \leq k-1,\  k\geq 4n.\nn 
\eeq 
Recall that the Riemann zeta function at the negative 
integers is given by Bernoulli numbers: $\zeta (-n)= -B_{n+1}/(n+1)$. 
 Using then the relation 
between these residua and the heat kernel coefficients 
\beq 
\mbox{Res}\, \zeta^\pm (3/2-k) &=&\mbox{Res}\, \sum_{n=0}^{k-1} A_{2n-1}^\pm 
(3/2-k)+ \mbox{Res}\, A_{2k-2}^\pm (3/2-k)+ \mbox{Res}\, A_{2k}^\pm (3/2-k) 
\nn \\ &=& \frac{C_k^\pm}{(4\pi)^{3/2} \Gamma (3/2 -k)}, 
\eeq 
we can calculate the $C_k^{tot}$, for which an explicit, closed expression 
can be given, in terms of generalized hypergeometric functions of the type 
of Eq. (\ref{23}) (what is quite clear from the consideration above).
The formula can be immediately turned into a computer code to perform 
the final part of the calculation. 
 It is important to observe that we have obtained, in fact, a closed, 
 explicit 
formula for the heat-kernel coefficients, even if the actual calculation 
is much more involved than that for the case of the Laplacian on the sphere 
with Robin boundary conditions. With the help of a fast computer we 
can construct (in a couple of hours)  a table of 
heat kernel coefficients $C_k^{tot}$ for integer values of $k$ up to 
 any (reasonable) order. The first few of them are given in Table 1. 
They are given  as power series on $a$ with rational coefficients. 
 
In a similar way we can proceed  with the calculation of the coefficients for 
half-integer index $C_{k+1/2}^{tot}$, $k\in\  \mbox{\bf N}$. 
The zeta functions $\zeta^\pm (s)$ have only one pole at integer $s$, 
 with an 
$a$-dependent residue. In fact 
\begin{equation} 
 \mbox{Res} A_1^{tot} (1)=\frac {r^2}{a^2 -1} 
\end{equation} 
 The values  
$\zeta^\pm (1-k)$ are needed and one has $M=2k$. 
Here  the $A_i^\pm (s)$ for $i$ odd, $i=2j-1$, 
$j\in\  \mbox{\bf N}_0$, do not contribute. 
The relevant  $A_i^{tot} (s)$ read now 
\beq 
A_0^{tot} (1-k)&=&\frac{r^{2-2k}}{2} \left[ \zeta (-2k) - 
\zeta (2-2k) \right], \nn \\ 
A_{0,a}^{tot} (1)&=&-\frac{r^2}{2} \left( 1 + \frac{\pi^2}{3}\right) 
\frac{ a^2}{1- a^2}, \nn \\ 
 A_{2k+1}^{tot} (1-k) &=& r^{2-2k}(-1)^k (k-1)! 
\sum_{j=0}^{3k+1} \left\{   
\frac{z^+_{2k+1,2j}}{(2k)!} \sum_{h=0}^\infty 
\frac{a^{2h} (2k+2h)!\, \Gamma (j+h+3/2)}{(2h)!\, \Gamma ( 
k+j+h+1/2)} \right.   \nn \\ 
&&\hspace{20mm} \left. 
-\frac{z^+_{2k+1,2j+1}}{(2k)!} \sum_{h=1}^\infty 
\frac {a^{2h-1}(2k+2h-1)!\, \Gamma (j+h+3/2)}{(2h-1)!\, 
\Gamma (k+j+h+1/2)} \right\}, \nn \\ 
 A_{2k-1}^{tot} (1-k) &=& -r^{2-2k}(-1)^k (k-1)! 
\sum_{j=0}^{3k-2} \left\{   
\frac{z^+_{2k-1,2j}}{(2k-2)!} \sum_{h=0}^\infty 
\frac{a^{2h} (2k+2h-2)!\, \Gamma (j+h+1/2)}{(2h)!\, \Gamma ( 
k+j+h-1/2)} \right.   \nn \\ 
&&\hspace{20mm} \left. 
-\frac{z^+_{2k-1,2j+1}}{(2k-2)!} \sum_{h=1}^\infty 
\frac {a^{2h-1}(2k+2h-3)!\, \Gamma (j+h+1/2)}{(2h-1)!\, 
\Gamma (k+j+h-1/2)} \right\}. 
\eeq 
And, for $n\in\  \mbox{\bf N}$, $n\leq k-1$, 
\beq 
A_{2n}^{tot} (1-k)&=&-4\, r^{2-2k}(k-1)! \left[ \zeta (2n-2k) - 
\zeta (2n-2k+2) \right]    
\nn  \\ 
&& \hspace{-31mm} \times \left\{ 
 \sum_{j=0}^{\mbox{min}(k-n-1,3n)}  \frac{z^\pm_{2n,2j}}{(2n-1)!} 
 \sum_{h=0}^{\mbox{min}(k-n-j-1,3n-j)} \hspace{-6mm} 
\frac{(-1)^{n+j+h} (2n+2h-1)!\,  a^{2h} }{ 
(2h)!\, (k-n-j-h-1)!\, (n+j+h-1)!} \right.  \\ && \left. 
\hspace{-31mm} -  
\sum_{j=0}^{\mbox{min}(k-n-2,3n-1)} \frac{ z^+_{2n,2j+1} 
}{(2n-1)!} \sum_{h=1}^{\mbox{min}(k-n-j-1,3n-j)} 
\hspace{-6mm} \frac{(-1)^{n+j+h} (2n+2h-1)!\,   a^{2h-1} }{ 
(2h-1)!\, (k-n-j-h-1)!\, (n+j+h-1)!} \right\}. \nn 
\eeq 
Notice again that the summation ranges are different depending now 
 on the fact that $k\leq 4n+1$ or $k \geq 4n+1$. 
The corresponding heat-kernel coefficients are readily calculated from 
\beq 
\zeta^\pm (1-k) =\sum_{n=0}^{k-1} A_{2n}^\pm (1-k) + A_{2k-1}^\pm (1-k) + 
A_{2k+1}^\pm (1-k) 
= \frac{(-1)^{k-1}(k-1)!}{(4\pi)^{\frac 3 2}}C_{k+\frac 1 2}^\pm.\nn 
\eeq 
 
A number of the coefficients $C_{k+\frac 1 2}^{tot} $are listed in Table 1 .  
If desired, the coefficients 
$a_n$ can be then obtained by using Eq. (\ref{defan}). Let us 
finish this section by noting  
that here we have calculated the $a$-dependent part of the 
full heat kernel expansion for the Euclidean Maxwell theory with 
the boundary conditions (\ref{bcon}) only. Complete result can be 
immediately 
recovered by taking $C^{tot}$ from Table 1, subtracting the $a^0$ part, 
and adding this to the known results \cite{vjmp1,MP,EspKam,elv2} 
for absolute boundary  
conditions. All coefficients, including $C_1$, are non-singular  
at $a=0$. 
 
\section{Conclusions} 
The conditions (\ref{bcon}) we have considered in this paper  
are mixed and contain tangential 
derivatives. For boundary conditions of this type only the 
first non-trivial coefficient $C_0$ was known analytically, 
up to now 
\cite{AE3}. Our result for $C_0$ agrees with the one in Ref. \cite{AE3}. 
The rest of the coefficients have been calculated here explicitly 
for the first time. Our calculations put restrictions 
on the universal functions (see \cite{AE-1}--\cite{DK2}) entering 
in higher heat kernel coefficients. 
Finally, we must point out that, even if we only list  
a few number of them in Table 1, in fact our program allows for the 
calculation of {\it any} coefficient to {\it any}  
degree of approximation in the 
$a$ dependence (notice that the coefficients are given by rational  
numbers, as it should be). However, to obtain each one of them as a  
closed function of $a$ is outside the scope of our purposes here.  
 
Several of the first heat kernel coefficients are singular at $a=1$. 
This is a manifiestation of the lack of strong ellipticity 
in the corresponding boundary value problem \cite{AE2,AE3}. 
 
A very interesting feature of our results is the vanishing of the 
$a$-dependent part of $C_{3/2}$ (or $a_2$), while the
heat kernel itself becomes singular at a certain (critical)
value of the coupling. This means that the 
conformal anomaly and the one--loop counterterms are {\it not} 
modified by the presence of the Chern--Simons boundary action, what 
 could be a reflection of a 
general property of Chern--Simons theories. Since
the fundamental explanation of very important physical 
phenomena of lower-dimensional QED, such as the fractional 
quantum Hall effect and high-temperature superconductivity, seem to
rely very heavily
on the fundamental structures provided by Chern--Simons theories,
we conclude that the property  we have here found (by looking to a 
particular model) could have important physical consequences for 
those subjects.
This issue is presently under investigation. 

\vspace{3mm} 
 
\noindent{\large \bf Acknowledgments} 
\ms 
 
EE is indebted with Andreas Wipf, Michael Bordag and specially Klaus 
Kirsten for enlightening 
discussions and with the members of the Institutes 
of Theoretical Physics of the Universities of Jena and Leipzig, 
where the main part of this work was done, for 
warm  hospitality. We thank the referee for constructive comments
that led to the improvement of the paper.  
 This investigation has been  supported by 
 CIRIT (Generalitat de Catalunya), by DGICYT (Spain), program PB96-0925, 
 and by the 
German-Spanish program Acciones Integradas, Ref. HA97-0053. 
DV thanks the Alexander von Humboldt Foundation and CRACENAS, 
grant 97-0-14.1-61, for financial support.  

\newpage 

\vspace*{10mm} 
 
\bce {\large \bf Table 1}: The first ten heat kernel coefficients.  
 \ece 
\beq 
C_{0} &=& 2\pi^2 r^3 \left[ (1-a^2)^{-3/2} -1\right] \nn \\ 
C_{1/2} &=& 8\pi^{3/2} (a^2-1)^{-1} \nn \\ 
C_{1} &=&  \pi^2 r \left[ \frac 1{2a^4} (1-a^2)^{-1/2} 
(2+a^2-8a^4) +\frac{117}{32} -\frac 1{a^2} -\frac 1{a^4} 
 \right] 
 \nn \\ 
C_{3/2} &=& \pi^{3/2} \frac{59}{45}  
  \nn \\ 
C_{2} &=& \frac{\pi^2}{r} \left( \frac{3}{ \pi} - \frac{1339}{4096} 
 + \frac{11\, a^2}{1024} + 
   \frac{115\, a^4}{4096} + \frac{435\, a^6}{16384} + \cdots 
  \right) 
  \nn \\ 
C_{5/2} &=& \frac{\pi^{3/2}}{r^2}  \left( \frac{117919}{45045} + 
   \frac{ 2048\, a^2}{15015} + 
   \frac{ 10112\, a^4}{109395}   + 
   \frac{ 816896\, a^6}{14549535} +  \cdots 
   \right) 
 \nn \\ 
C_{3} &=&  \frac{\pi^2}{r^3}  \left( \frac{ 
21 }{8\pi } -  \frac{57455}{ 393216} + 
   \frac{ 6787\, a^2}{65536 } + \frac{ 43053\, a^4}{ 1048576} + 
   \frac{ 14431\, a^6}{ 1048576} +  \cdots  \right) \nn \\ 
C_{7/2} &=& \frac{\pi^{3/2}}{r^4}  \left( \frac{ 
371148101 }{116396280} + 
   \frac{ 571904\, a^2 }{2909907}  + 
   \frac{ 1259008\, a^4 }{37182145} - 
   \frac{ 310784\, a^6}{42902475} +  \cdots 
 \right) 
  \nn \\ 
C_{4} &=&  \frac{\pi^2}{r^5}  \left( \frac{ 
633}{160 \pi} - \frac{ 16417555 }{201326592} + 
   \frac{ 468537\, a^2 }{4194304} - 
   \frac{ 64091\, a^4 }{16777216} - 
   \frac{ 515873\, a^6 }{33554432} +  \cdots  \right) 
   \nn \\ 
C_{9/2} &=&  \frac{\pi^{3/2}}{r^6}  \left( \frac{ 
 399265868279}{60235074900} + 
   \frac{ 997830656\, a^2 }{5019589575 }  - 
   \frac{ 386748416\, a^4 }{8562829275 
    }\right.  \nn \\ && \left. \hspace*{12mm} - 
   \frac{ 68230270976\, a^6 }{1933976154825} + 
 \cdots \right)  \nn 
\eeq 
\end{document}